\documentclass[authoryear,preprint,review,12pt]{elsarticle}

\usepackage{graphics}

\usepackage{graphicx}

\usepackage{epsfig}

\usepackage{amssymb}

\usepackage{amsthm}

\usepackage{lineno}

\journal{New Astronomy}

\begin{document}

\begin{frontmatter}



\title{Brightness variations in totally-eclipsing binary GSC\,4589-2999}


\author{E. Sipahi\corref{cor1}}
\ead{esin.sipahi@mail.ege.edu.tr}
\cortext[cor1]{Corresponding author}

\author{H. A. Dal}

\author{O. \"{O}zdarcan}

\address{Ege University, Science Faculty, Department of Astronomy and Space Sciences, 35100 Bornova, \.{I}zmir, Turkey}

\begin{abstract}
We present multi-colour CCD photometry of GSC\,4589-2999 obtained in 2008 and 2009. The observations indicate that the system is an active Algol binary. Based on the new data, the mean brightness of the system is decreasing through the years 2007-2009. The light curves obtained in 2008-2009 are modelled using the Wilson-Devinney code. We also discussed the light and colour variations of the system at different orbital phases. Evidence suggests that these brightness and colour variations are due to the rotation of unevenly distributed starspots on two components of the system.
\end{abstract}

\begin{keyword}

techniques: photometric - (stars:) binaries: eclipsing  -  stars: individual: (GSC\,4589-2999)

\end{keyword}

\end{frontmatter}


\section{Introduction}\label{S1}

GSC\,4589-2999 ($V$=10$^{m}$.54) is classified as an eclipsing binary star in the SIMBAD database. The system was discovered by \citet{Lia07}, who provided an R band light curve. The binary has been included in our multi-colour photometric programme. \citet{Lia11} presented BVRI light curves and the radial velocities. The physical properties and the absolute parameters of the components were given in their study.

We report the new photometric observations, and discussed the light and colour variations of the system. The system is important not only because it joins the group of Algol type binary systems, but also because it is a part of an active Algol systems with total eclipse as we have showed in this study.

\section{Observations}\label{S2}

Observations were acquired with a thermoelectrically cooled ALTA U+42 2048$\times$2048-pixel CCD camera attached to a 40-cm Schmidt-Cassegrain MEADE telescope at Ege University Observatory. The $BVRI$-band observations were recorded over twelve nights in 2008 and fifteen nights in 2009. We used GSC\,4589-2984 and GSC\,4589-2842 as a comparison and check star, respectively. The comparison and check stars used in all the observations are the same stars used by \citet{Lia11}. There was no variation observed in the brightnesses of the comparison star.

During the observations, we obtained four primary and one secondary times of minimum light. All the available times of minima are given in Table~\ref{T1}. We combined our times of minima with the literature and obtained the new light elements as follows:

\begin{equation}
T{\rm (HJD)} = 2455103.4986(19) + 1^{d}.688646(6) \ \times \ E \ .\label{E1}
\end{equation}

All the phases in the tables and the figures are calculated with these new light elements. The $O-C$ diagram constructed by using the light elements given in Equation~\ref{E1} indicates no variability of the period.

\begin{table}[!htb]
\caption{The times of minimum light for GSC\,4589-2999}\label{T1}
\centering
\begin{tabular}{ccccccc}
\hline\hline
O	&	Year	& $(O-C)_{I}$	&	$(O-C)_{II}$	&	Type	&	Filter	&	REF	\\
\hline
54290.4167	&	2007.51	&	0.0653	&	0.0012	&	II	&	$RI$	&	1	\\
54306.4601	&	2007.56	&	0.0653	&	0.0024	&	I	&	$BVRI$	&	1	\\
54312.3603	&	2007.57	&	0.0548	&	-0.0076	&	II	&	$BVRI$	&	1	\\
54377.3822	&	2007.75	&	0.0587	&	0.0014	&	I	&	$R$	&	1	\\
54399.3355	&	2007.81	&	0.0579	&	0.0023	&	I	&	$R$	&	1	\\
54637.4327	&	2008.46	&	0.0373	&	0.0004	&	I	&	$BVRI$	&	1	\\
54642.4998	&	2008.48	&	0.0381	&	0.0016	&	I	&	$V$	&	1	\\
54691.4703	&	2008.61	&	0.0340	&	0.0013	&	I	&	$V$	&	1	\\
54752.2629	&	2008.78	&	0.0306	&	0.0027	&	I	&	$BVRI$	&	2	\\
54833.3082	&	2009.00	&	0.0145	&	-0.0070	&	I	&	$BVRI$	&	2	\\
55055.3703	&	2009.61	&	0.0023	&	-0.0019	&	II	&	$I$	&	2	\\
55076.4838	&	2009.66	&	0.0061	&	0.0036	&	I	&	$BVRI$	&	3	\\
55103.4982	&	2009.74	&	0.0000	&	-0.0004	&	I	&	$BVRI$	&	3	\\
\hline
\end{tabular}
\begin{list}{}{}															
\item[1]{\small \citet{Lia09}}
\item[2]{\small \citet{Sip09}}
\item[3]{\small This Study}																													
\end{list}
\end{table}

\begin{figure}[!ht]
\hspace{0.5cm}
\includegraphics[width=\textwidth]{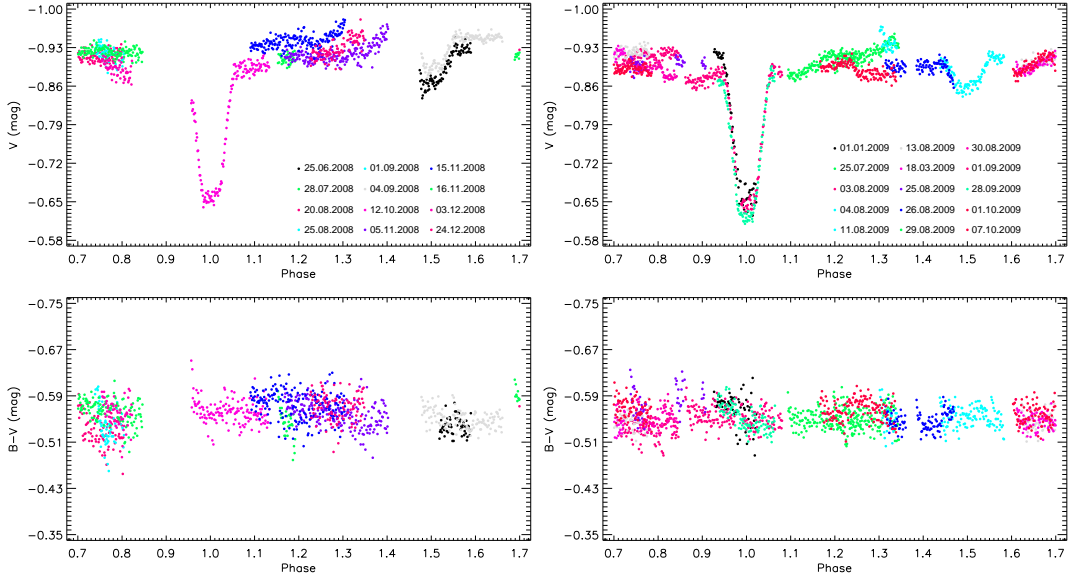}
\caption{2008 (left) and 2009 (right) light and colour curves obtained in this study.}\label{F1}
\end{figure}

The observed V light and B-V colour curves are shown in Figure 1. The shape of the light curves indicate that GSC\,4589-2999 is an Algol type binary. The system has a totality in the secondary eclipse, which lasts about $\sim$1.97 hours. The light curve variations which obtained in 2008 and 2009 are the different. Both minima are asymmetrical enough in the light curves. The mean depths of eclipses in B, V, R, and I filters are 0$^{m}$.252, 0$^{m}$.236, 0$^{m}$.222, and 0$^{m}$.204 in the primary minimum and 0$^{m}$.056, 0$^{m}$.061, 0$^{m}$.066, and 0$^{m}$.083 in the secondary minimum, respectively. The light curves also display an asymmetrical light variations at outside of the eclipse. The amplitudes of these variations are $\sim$0$^{m}$.052, 0$^{m}$.059, 0$^{m}$.062, 0$^{m}$.069 in the $B$, $V$, $R$, and $I$ filters, respectively. These light variations are very similar to the photometric variations commonly observed for RS CVn type variables, which appear to arise from an uneven distribution of star spots over the surface of the rotating star \citep{Hal76}. The general features are not similar to those of the light curves previously taken from \citet{Lia11}.

\section{Discussion and conclusions}\label{S3}

\subsection{The activity of both components}

The previous light curves of the system have been presented and their variations discussed by \citet{Lia11}. They argued for a cool spot in the secondary component. However, all the phenomena observed in the light curves could not be explained by taking into account only one component's spot activity. GSC\,4589-2999 has a totality in the secondary eclipse. The brightness level at totality of the secondary minimum change in each observation. This light variability seen in the total secondary eclipse should be originated from the more massive primary star. \citet{Lia11} analysed the light curves with the radial velocity curve and derived the orbital parameters of the system. Using these parameters of the system in this study, the R-band light curves observed in 2008 and 2009 were modelled by fitting dark spots onto the surface of the primary and secondary components with using the PHOEBE V.0.31a software \citep{Prs05}. The result spot parameters are listed in Table~\ref{T2}. The synthetic light curves obtained from the modelled light curves are seen in Figure 2.

\begin{figure}[!ht]
\hspace{2.9cm}
\includegraphics[width=17cm]{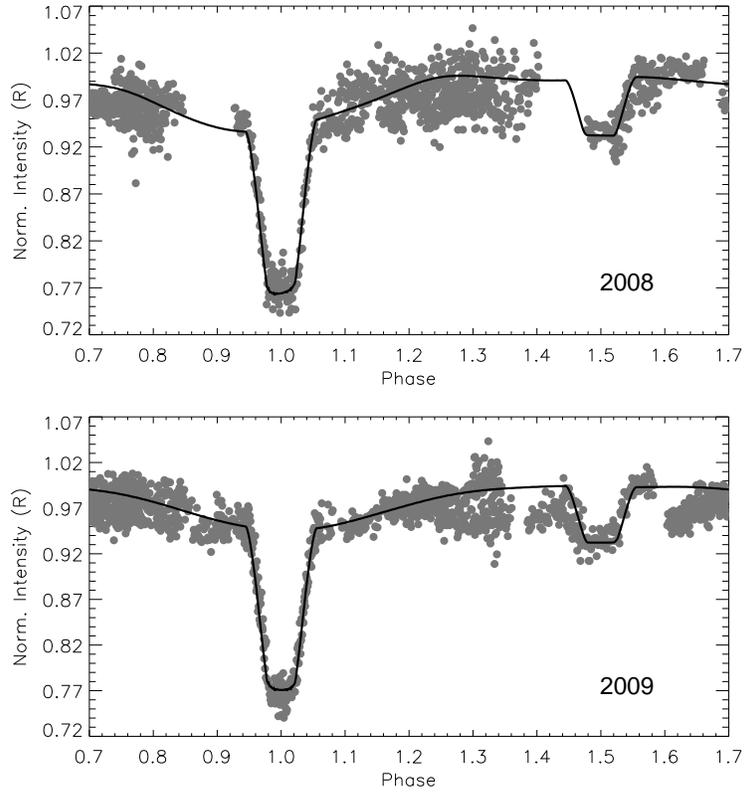}
\vspace{0.1cm}
\caption{The R-band light curves observed in 2008 (upper panel) and 2009 (bottom panel) and the synthetic curves (lines) derived from the light curve models.}\label{F2}
\end{figure}

\begin{table}[!htb]
\caption{The spot parameters of the components in 2008 and 2009.}\label{T2}
\centering
\begin{tabular}{clcccc}
\hline\hline
Year 	&	 Parameter 	&	\multicolumn{2}{c}{Primary}			&	\multicolumn{2}{c}{Secondary}			 \\
&		&	Spot I	&	Spot II	&	Spot I	&	Spot II	 \\
\hline															
2008	&	 $Co-Lat$ ($^{\circ}$) 	&	56	&	40	&	66	&	 - \\
 	&	 $Co-Long$ ($^{\circ}$) 	&	0	&	70	&	70	&	 - \\
 	&	 $R$ ($^{\circ}$) 	&	15	&	15	&	25	&	 - \\
 	&	 $T_{Spot}$/$T_{Sur}$ 	&	0.7	&	0.8	&	0.6	&	 - \\
2009	&	 $Co-Lat$ ($^{\circ}$) 	&	20	&	40	&	86	&	 85 \\
 	&	 $Co-Long$ ($^{\circ}$) 	&	350	&	45	&	290	&	 95 \\
 	&	 $R$ ($^{\circ}$) 	&	20	&	10	&	20	&	 25 \\
 	&	 $T_{Spot}$/$T_{Sur}$ 	&	0.6	&	0.8	&	0.6	&	 0.60 \\
\hline
\end{tabular}
\end{table}

\subsection{The shape of the light curves at the bottom of the primary minima}

\begin{figure}[!ht]
\hspace{1.5cm}
\includegraphics[width=15cm]{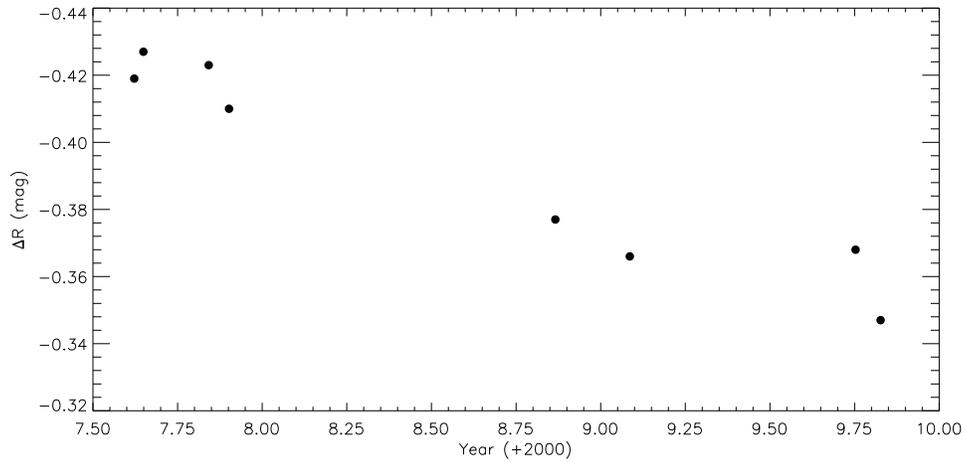}
\vspace{0.1cm}
\caption{The variation of the brightness level of the primary minimum of GSC\,4589-2999.}\label{F3}
\end{figure}

The system shows at primary minimum either a curved bottom normal for a partial eclipse or a flat bottom which resembles a total eclipse. In order to study further changes at the primary minima, the observations in $R$ filter are collected. The type of the bottom of the primary eclipse October 12, 2008 was total, later, on  September 28, 2009 it was partial. When a minimum showed asymmetry, the bottom of the eclipse shows partial shape. Both bottom types (partial or flat) of the brightness level at the primary minima have almost different. The shape of the minima in the light curves will be changed whether the spots are located at minimum phases toward to observing side or not. The brightness level of the primary minima in all filters are listed in Table~\ref{T3}. The variation of the brightness level in R filter are seen in Figure~\ref{F3}. As seen from Table~\ref{T3} and Figure~\ref{F3}, the brightness level of the primary minimum is decreasing $\sim$0$^{m}$.09 during three years. It seems likely that the brightness changes in the secondary component caused by spots is responsible for these variations.    

We conclude here that the system exhibits both types of bottoms of the primary minimum and the change between two types occurs rather frequently and rapidly. Many investigators have reported such activities on active Algol binaries. \citet{Ols82a, Ols82b} discovered the variable depth, which was only a little but considered real, of the total primary minimum in U\,Sge. \citet{Ols87} proposed an explanation that the brightness changes could be due to the temperature variations on the photosphere of the cool component.

\subsection{The out-of-eclipse variation}

The brightness of the system is decreasing though the years 2007-2009. The observations of \citet{Lia07} showed that the light variations of out-of-eclipse had the smallest amplitude observed in 2007. In the observing season 2008, the amplitude increased to 0$^{m}$.024. These light variations are caused by the effects of cool spot activity existing on both components of the system. 

In addition, we noticed considerable changes in the shape of the light curve on timescales of a few weeks in our observations. The spot distribution on the surfaces of the components are rapidly changing. The effectivity of the spots vary from one month to the next with an amplitude of approximately 0$^{m}$.07 in R filter. A similar behaviour was found in the light variation of DK\,CVn by \citet{Dal12}.    

Examining carefully the light variation between the phases 0.70-0.95 in Figure 1, we revealed a rapid light variation at the same phases. One can notice that the observations made in August 03, 13, 18 and 25 revealed that the levels of light follow themselves in the observing season 2009. The light variations obtained in August 30, September 1 and October 1 also follow themselves in the general light variation. Although both these groups are corresponding to the same phases between 0.70 and 0.95, their levels are different from each other.

\begin{table}[!htb]
\caption{The brightness level of the primary minima of GSC 4589-2999}\label{T3}
\begin{center}
\begin{tabular}{cccccc}
\hline\hline
HJD	&	B	&	V	&	R	&	I	&	Type	\\
(+24 00000)	&	(mag)	&	(mag)	&	(mag)	&	(mag)	&		\\
\hline															
54296.3289	&		&		&	-0.419	&		&	I	\\
54306.4590	&		&		&	-0.427	&		&	I	\\
54377.3825	&		&		&	-0.423	&		&	I	\\
54399.3345	&		&		&	-0.410	&		&	I	\\
54752.2603	&	-1.209	&	-0.660	&	-0.377	&	-0.096	&	I	\\
54833.3150	&	-1.212	&	-0.656	&	-0.366	&	-0.106	&	I	\\
55076.4784	&	-1.175	&	-0.644	&	-0.368	&	-0.092	&	I	\\
55103.4985	&	-1.153	&	-0.614	&	-0.347	&	-0.065	&	I	\\
54290.4179	&		&		&	-0.586	&		&	II	\\
54643.3433	&		&	-0.860	&	-0.569	&	-0.266	&	II	\\
54714.2662	&	-1.450	&	-0.894	&	-0.601	&	-0.289	&	II	\\
55055.3707	&	-1.411	&	-0.856	&	-0.574	&	-0.265	&	II	\\
\hline
\end{tabular}
\end{center}
\end{table}

\subsection{The short-term variation behaviour in light and colour curves}

Apart from the long-term variations in the light curves, examining each nightly observation day by day demonstrated that there are also some distinctive short-term variations in even one observing season. As seen from Figure 1, there are some distinctive relations between light and colour curves. Comparisons of the observations phased between the same range indicate that the brightness level of the system is rapidly changing in even one season. In observing season 2008, the observations obtained in August 20, November 16, and December 03 are corresponding to the same phases between 0.7 and 0.8, however their brightness levels are different from each other. In 2008, the brightness level in the $V$-band decreased from August 20 to December 03, while the $B-V$ colour gotten bluer. Similarly, the observations in July 28, October 12, and December 15 are corresponding to the same phase range. The levels of $V$-band magnitude and $B-V$ colour observed in July 28 and October 12 are almost the same. However, the brightness level in the $V$-band increased in December 15, and the $B-V$ colour gotten bluer. In the same way, the level of the secondary minima is also changing. The observations in June 25 and September 04 are corresponding to the same phase interval between 0.46 and 0.68. Comparison of these observations indicates that the level of the secondary minimum is changing. As seen in Figure 1, from June 25 to September 04, the level of the $V$-band magnitude increased, while the $B-V$ colour gotten bluer. In observing season 2009, there are six observing nights, in which the observations are corresponding to the same phases between 0.6 and 0.8. The observations made in August 03, 13 and 18, revealed that the levels of both light and colour are almost the same. Similarly, the levels observed in August 30, September 01 and October 01 are the same among themselves. On the other hand, the mean levels obtained from these two observing groups are clearly different from each other. The level of the $V$-band magnitude decreased from the previous three night to the latter, while the $B-V$ colour gotten bluer in contrast to the $V$ light. The same phenomenon is seen in the observations made in July 23, August 04, 29 and October 07, which all of them are corresponding to the same phases between 0.10 and 0.35. The levels of the light and colour are almost the same in July 23, August 04, 29. However, the $V$-band magnitude decreased in October 07, while the $B-V$ colour gotten bluer.

\section{Results}

Based on the results presented in the previous sections we draw conclusions as follows;

$\bullet$ The main features of the light curves are: a) the asymmetry in the minima and the unequal maxima; b) the larger brightness variations at total primary minima than at any other phase. The observed light variations of GSC\,4589-2999 can be well explained by the spot activity occurring on both components of the system. 
	
$\bullet$ When the spots located on the components are seen at minimum phases, the amplitude of the light variation is larger than in other phases. These phenomenon is generally referred to a greater concentration of active regions when the active longitude is on the hemisphere facing the companion \citep{Lan98}. 
	
$\bullet$ The out-of-eclipse brightness variation of the system were examined in detailed. During the three years the average brightness has decreased about 0$^{m}$.024. The larger brightness decreases occur at the primary eclipse. This indicates a change in the spots on the secondary component, that may be due to an increase in the spot coverage. The amplitude of each seasonal variation indicates the degree of spottedness of the stellar surface. The light variations during out-of-eclipses is highly variable, changing at times within a few week. Similar results have found for DK\,CVn \citep{Dal12}. 
	
$\bullet$ Another important feature observed in the system is colour variation behaviour. The $B-V$ colour is sometimes getting bluer, while the brightness of the system is increasing. However, the $B-V$ colour is also getting bluer, while the brightness level is decreasing. As it is well known, the long-term photometric studies of Solar like old stars demonstrated that the bright-hot features such as faculae provide larger excess to the total light, instead of the cool-dark features such stellar spots. These bright-hot features are generally located around the cool spots in the active regions on the stellar surface. In such a case, the cool-dark spots cause general light variations due to the rotational modulation. In addition, the bright-hot faculae cause some additional blue excess in $B-V$ and especially $U-B$ \citep{Wil94, Ber05}. If GSC\,4589-2999 is a Solar like old system as offered by \citet{Lia11}, the bright-hot features like faculae could occur on the stellar surface and also can cause the variations seen in the colour curves of both 2008 and 2009.
	
$\bullet$ The short-term variations such as those observed from the system of GSC\,4589-2999 are generally observed in the young and rapidly rotating BY Dra type stars \citep{Str09}, such as AB\,Dor \citep{Ama01} and LO\,Peg \citep{Pan11} and also some low-mass close binary systems, such as DK\,CVn \citep{Dal12}. However, according to the results obtained by \citet{Lia11}, GSC\,4589-2999 is an evolved system. In this case, the rapid variability seen in the system can not be caused due to young age, but it should be due to the binarity effect. 

$\bullet$ It is clear that more observations will be necessary to describe the long-term activity of GSC\,4589-2999. Similar variations have also been observed for other active Algol binaries (Olson, 1987). We conclude that GSC\,4589-2999 is a new Algol type active binary with a changing light curve.

$\bullet$ GSC\,4589-2999 has very high activity level. In terms of stellar astrophysics, the chromospherically active binary systems like GSC\,4589-2999 are very important to understand the stellar evolution and also evolution of the angular momentum. Firstly, developing a model about the formation of the close binaries, \citet{Roc02} indicate some unexpected systems, which are chromospherically young without high Li abundance, while they are kinematically old. According to them, their components are generally rapidly rotating due to the binarity. The initial parameters of a system are important for such a model. The parameters of the system, which was in the main-sequence stage, are generally assumed as the initial parameters. In this respect, the chromospherically active binary systems, whose components are still in the main-sequence, become very important to determine the absolute parameters such as mass, radii, rotational velocity, and so the stellar angular momentum in this stage. In addition, as it is seen from the literature, there are no more chromospherically active systems, enough to determine general outline of their absolute parameters in the main-sequence stage. Secondly, there is another unsolved problem about the stellar absolute parameters for the chromospherically active components of the binaries. Several studies, such as \citet{Lop07}, \citet{Cas08}, \citet{Mor08, Mor10}, \citet{Fer09}, \citet{Tor10}, and \citet{Kra11}, have already revealed that the radii of chromospherically active components are found dramatically larger than expected ones. The main effect on the larger radius seems to be the chromospheric activity. Although the general dividing is seen for the low mass stars, there is no information for the stars from the spectral types G and K due to fewer samples. In this respect, so many systems like GSC\,4589-2999 should be found to multiple the number of samples. In addition, \citet{Lop07}, \citet{Mor08} and \citet{Mor10} mentioned that the best way to find radius of a star is the light curve analyses of eclipsing binaries, especially having minima with totality.

Finally, it should be noted that the photometric observations of GSC\,4589-2999 are continuing. Long-term photometry of the system will help paint a clearer picture of the activity nature of this interesting active binary.

\section*{Acknowledgments} The authors acknowledge generous allotments of observing time at the Ege University Observatory. We also wish to thank the referee for useful comments that have contributed to the improvement of the paper.


\begin{thebibliography}{23}


\bibitem[Amado et al.(2001)]{Ama01} Amado, P.J., Cutispoto, G., Lanza, A.F., Rodon\'{o}, M., 2001, AB Doradus: Long and Short Term Light Variations and Spot Parameters, in \emph{Cool Stars, Stellar Systems, and the Sun (11th Cambridge Workshop)}, (Eds.) Garc\'{i}a L\'{o}pez, R.J., Rebolo, R., Zapaterio Osorio, M.R., Proceedings of a meeting held at Puerto de la Cruz, Tenerife, Spain, 4–8 October 1999, vol. 223 of ASP Conference Series, pp. 895–900, Astronomical Society of the Pacific, San Francisco, U.S.A. 11, 6.1, 5
\bibitem[Berdyugina(2005)]{Ber05} Berdyugina, S.V., 2005, LRSP, 2, 8 (P.36)
\bibitem[Casagrande et al.(2008)]{Cas08} Casagrande, L., Flynn, C., and Bessell, M., 2008, MNRAS, 389, 585
\bibitem[Dal et al.(2012)]{Dal12} Dal, H.A., Sipahi, E., \"{O}zdarcan, O., 2012, PASA, 29, 54
\bibitem[Fernandez et al.(2009)]{Fer09} Fernandez, J.M., et al., 2009, ApJ, 701, 764
\bibitem[Hall(1976)]{Hal76} Hall, D.S., 1976, in "Multiple Periodic Phenomena in Variable Stars", IAU Colloq. No. 29, Part I, edited by W.S. Fitch (Reidel, Dordrecht)
\bibitem[Kraus et al.(2011)]{Kra11} Kraus, A.L., Tucker, R.A., Thompson, M.I., Craine, E.R., Hillenbrand, L.A., 2011, ApJ, 728, 48
\bibitem[Lanza et al.(1998)]{Lan98} Lanza, A. F., Catalano, S., Cutispoto, G., Pagano, I.,  Rodon\'{o}, M., 1998, A\&A, 332, 541
\bibitem[Liakos and Niarchos(2007)]{Lia07} Liakos, A., Niarchos, P., 2007, IBVS, 5900, 2
\bibitem[Liakos and Niarchos(2009)]{Lia09} Liakos, A., Niarchos, P., 2009, IBVS, 5897, 1
\bibitem[Liakos et al.(2011)]{Lia11} Liakos, A., Bonfini, P., Niarchos, P., Hatzidimitriou, D., 2011, AN, 332, 602
\bibitem[L\'{o}pez-Morales(2007)]{Lop07} L\'{o}pez-Morales, M., 2007, ApJ, 660, 732
\bibitem[Morales et al.(2008)]{Mor08} Morales, J.C., Ribas, I., Jordi, C., 2008, A\&A, 478, 507
\bibitem[Morales et al.(2010)]{Mor10} Morales, J.C., Gallardo, J., Ribas, I., Jordi, C., Baraffe, I., Chabrier, G., 2010, ApJ, 718, 502
\bibitem[Olson(1982a)]{Ols82a} Olson, E.C., 1982a, ApJ, 259, 702
\bibitem[Olson(1982b)]{Ols82b} Olson, E.C., 1982b, PASP, 94, 700
\bibitem[Olson(1987)]{Ols87} Olson, E.C., 1987, AJ, 94, 1043
\bibitem[Pandey et al.(2011)]{Pan11} Pandey, J.C., Medhi, B.J., Sagar, R., 2011, IAUS, 273, 455
\bibitem[Pr\v{s}a \& Zwitter, 2005]{Prs05} Pr\v{s}a, A., Zwitter, T., 2005, ApJ, 628, 426
\bibitem[Rocha-Pinto et al.(2002)]{Roc02} Rocha-Pinto, H.J., Castilho, B.V., Maciel, W.J., 2002, A\&A, 384, 912
\bibitem[Sipahi et al.(2009)]{Sip09} Sipahi, E., Dal, H.A., \"{O}zdarcan, O., 2009, IBVS, 5904, 1
\bibitem[Strassmeier(2009)]{Str09} Strassmeier, K.G., 2009, A\&ARv, 17, 251
\bibitem[Torres et al.(2010)]{Tor10} Torres, G., Andersen, J., Gim\'{e}nez, A., 2010, A\&ARv, 18, 67
\bibitem[Wilson(1994)]{Wil94} Wilson, P.R., 1994, "Solar and stellar activity cycles", First Edition, ed. R.F. Carswell, D.N.C. Lin and J.E. Pringle, United States of America by Cambridge University Press, New York, p.118
\end{thebibliography}
\end{document}